\documentclass[%
 reprint,
superscriptaddress,
 amsmath,amssymb,
 aps,
]{revtex4-1}

\usepackage{graphicx}
\usepackage{dcolumn}
\usepackage{bm}

\begin{document}
\preprint{APS/123-QED}
\title{Evidence of hyperdimensional topological defects \\ in a ferroelectric supercrystal phase-transition}

\author{Feifei Xin}
\affiliation{Dipartimento di Fisica, Universit\`{a} di Roma ``La Sapienza'', 00185 Rome, Italy}
\affiliation{College of Physics and Materials Science, Tianjin Normal University, 300387, Tianjin, China}

\author{Fabrizio Di Mei}
\affiliation{Dipartimento di Fisica, Universit\`{a} di Roma ``La Sapienza'', 00185 Rome, Italy}

\author{Ludovica Falsi}
\affiliation{Dipartimento di Fisica, Universit\`{a} di Roma ``La Sapienza'', 00185 Rome, Italy}
\affiliation{Dipartimento S.B.A.I., Sezione di Fisica, Universit\`{a} di Roma "La Sapienza" ,00161 Rome, Italy}

\author{Davide Pierangeli}
\affiliation{Dipartimento di Fisica, Universit\`{a} di Roma ``La Sapienza'', 00185 Rome, Italy}

\author{Galina Perepelitsa}
\affiliation{The Department of Applied Physics, The Hebrew University of Jerusalem, Jerusalem 9190401,Israel}

\author{Yehudit Garcia}
\affiliation{The Department of Applied Physics, The Hebrew University of Jerusalem, Jerusalem 9190401,Israel}

\author{Aharon J. Agranat}
\affiliation{The Department of Applied Physics, The Hebrew University of Jerusalem, Jerusalem 9190401,Israel}

\author{Eugenio DelRe}
\affiliation{Dipartimento di Fisica, Universit\`{a} di Roma ``La Sapienza'', 00185 Rome, Italy}
\affiliation{ISC-CNR, Universit\`a di Roma ``La Sapienza'', 00185 Rome, Italy}

\date{\today}

\begin{abstract}
\noindent We perform real-time stereoscopic wide-area imaging of a ferroelectric phase-transition in KTN:Li.  Spontaneous polarization is observed to form a thermally hysteretic 3D lattice of mutually interlinked closed-flux patterns that spans the entire sample.     Results are compatible with a supercrystal of $N=4$ topological texture defects arising as the three-fold spatial and one-fold time-inversion symmetries are simultaneously broken. Each lattice site of the texture supercrystal emerges as the projection in actual space of an S$^3$ hypersphere, an extended volume Hopf-link fabric able to screen both volume charge and ferroelectric strain.

\end{abstract}
\pacs{Valid PACS appear here}

\maketitle 


The spontaneous breaking of symmetry is one of the most fundamental concepts in modern physics \cite{Strocchi2008}.  Contemplating the emergence of new physical features, such as charge, polarization, magnetization, and mass, from an unstable original vacuum state at a specific critical temperature, it provides the basic model to unify different fields through a single general explanatory picture. Symmetry breaking also predicts specific accompanying consequences, principle among these are the so-called topological defects \cite{Kibble1976,Vilenkin1994}.   According to the Kibble mechanism, these can naturally form at the critical temperature $T_C$ on consequence of the combined effect of equilibrium thermal fluctuations and the spatial extension of the system \cite{Griffin2012,Zong2018}.  Here the defects are domain-mismatch singularities, regions of high-symmetry that populate the otherwise lower symmetry environment \cite{Bunkov2000,Lin2014, Kosterlitz2017}.  

\begin{figure}[t]
\centering
\includegraphics[width=1.0\columnwidth]{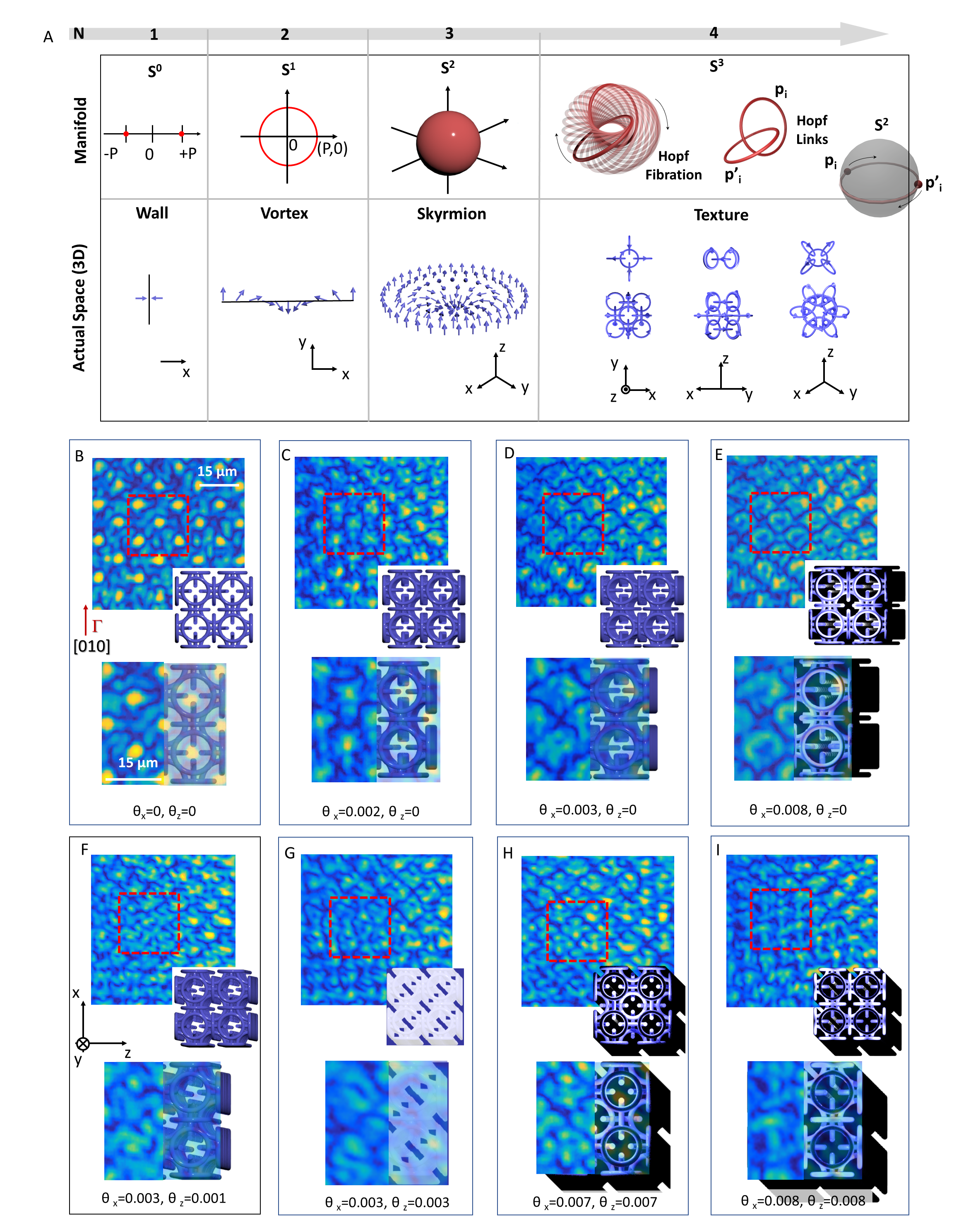}
\caption{(A) Dimensional hierarchy of topological defects when $N$ scalar fields suffer simultaneous symmetry-breaking (see text). Vacuum state  manifold (red region, left colum) and order-parameter polarization vector $\mathbf{P}$ (blue arrows) in actual space (right column). (B-I) Stereoscopic transmission imaging of KTN:Li cooled through its phase transition (to $T=T_C-3$K) and evidence of a texture supercrystal.  Each panel reports (top row) the imaging of the $xz$ output facet and (bottom) comparison to the model polarization texture distribution (inset, model), for different rotations. A real-time exploratory scan of the texture can be found in the Supplemental Video.}
\label{figure1}
\end{figure}

Consider a phase-transition that involves the breaking of symmetry of an $N$-component scalar field. As illustrated in Fig. 1A, the topology and invariants of the defect are determined by the properties of the ($N-1$)-dimensional sphere, $S^{N-1}$, that forms the associated degenerate vacuum manifold \cite{Kibble1976}. For $N=1$, the vacuum manifold is $S^0$, i.e., a pair of points on a line.  The order parameter that flags the degenerate vacuum states is then a one  component vector $\mathbf{P}$ that flips on crossing the defect ($\mathbf{P} \rightarrow -\mathbf{P}$), leading to the domain walls observed, for example, in ferromagnets (first column in Fig. 1A) \cite{Tetienne2015}.  For $N=2$ the defect appears as a vortex, $\mathbf{P}$ is on a plane and topological defects are associated to an $S^1$ manifold (a circle, second column of Fig. 1A) \cite{Naumov2004,Yadav2016,Hong2017,Hsu2019}. When $N=3$, topological invariants emerge as the vector field $\mathbf{P}$ is mapped onto the sphere $S^2$, a structure that is called a skyrmion (third column in Fig. 1A) \cite{Nagaosa2013,Posnjak2017,Yu2018,Kanazawa2020,Das2019,Das2020}. When, at one given critical temperature, a four-fold $N=4$ inversion symmetry-breaking occurs, then the resulting 4-component order parameter $\mathbf{P}$ is mapped onto a hypersphere $S^3$ manifold, a topological defect whose manifestation is termed a texture (fourth column in Fig. 1A) \cite{Kibble1976,Note1}. The hypersphere can be represented in 3D space using the so-called Hopf fibration, a series of interlinked rings (the Hopf links), each ring corresponding to a point $p_i$ on the standard sphere ($S^2$ in Fig. 1A top row, fourth column).   The texture may then appear in actual 3D space as a Hopf-linked structure, while an extended texture supercrystal  may form as a 3D lattice of interlinked Hopf-links (fourth column, second row in Fig. 1A).  Furthermore, since $N=4$ textures represent nonsingular unstable topological solitons in purely Hamiltonian systems, they will form stable defects when accompanied by the breaking of time-inversion symmetry   \cite{Derrick1964}. To date, however, no observation in 3D actual space of a Kibble hyperdimensional texture has been reported.

In this paper we provide, for the first time, evidence of a hyperdimensional Kibble texture.  The texture is found as a KTN:Li crystal undergoes a ferroelectric phase-transition. Spontaneous polarization is observed to form Hopf-linked closed-loop topological defects that extend into a volume periodic 3D supercrystal.  In agreement with the predictions of the Kibble mechanism, the volume Hopf-link pattern is characterized by strong thermal hysteresis, demonstrating that the extra wrapped-up dimension is in fact the product of time-inversion symmetry breaking.   


\begin{figure*}[t]
\centering
\includegraphics[width=1.6\columnwidth]{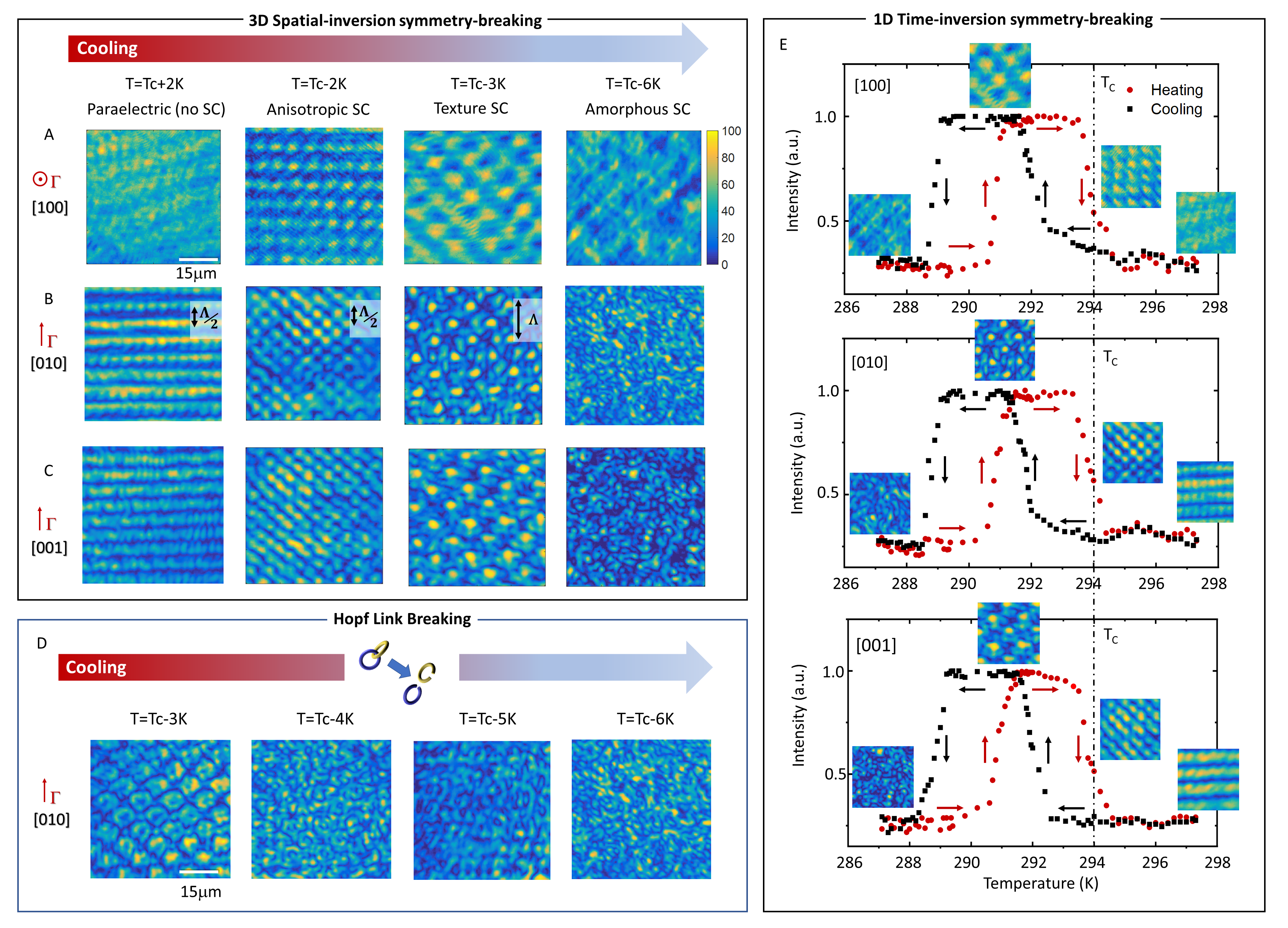}
\caption{Evidence of 3D-space-plus-1D-time inversion symmetry breaking. (A-C) Space-inversion symmetry analysis. Image of the (A)  yz facet, (B) xz facet, and (C) xy facet, for $T=T_C+2$K (first column),  $T=T_C-2$K (second column), $T=T_C-3$K (third column), $T=T_C-6$K (fourth column). $\Lambda$ is the lattice constant of the texture SC, while $\Lambda/2$ is the lattice constant of the Anisotropic SC.   Far-field analysis of the texture SC for the different temperatures is reported in Fig. S2 in the Supplemental Material. (D) Breaking of Hopf links as the sample is cooled as observed imaging the $xz$ facet (in the same conditions of Fig. 1 E). (E) Time-inversion symmetry analysis. Thermal hysteresis in the texture SC lattice structure for the (top) yz (pulling/growth $\Gamma$ direction), (center) xz, and (bottom) xy facet. Black squares and red circles report the intensity of the first and second order diffraction peaks of the far-field images of the SC on cooling and heating, respectively.}
\label{figure2}
\end{figure*}

\begin{figure*}[t]
\centering
\includegraphics[width=1.8\columnwidth]{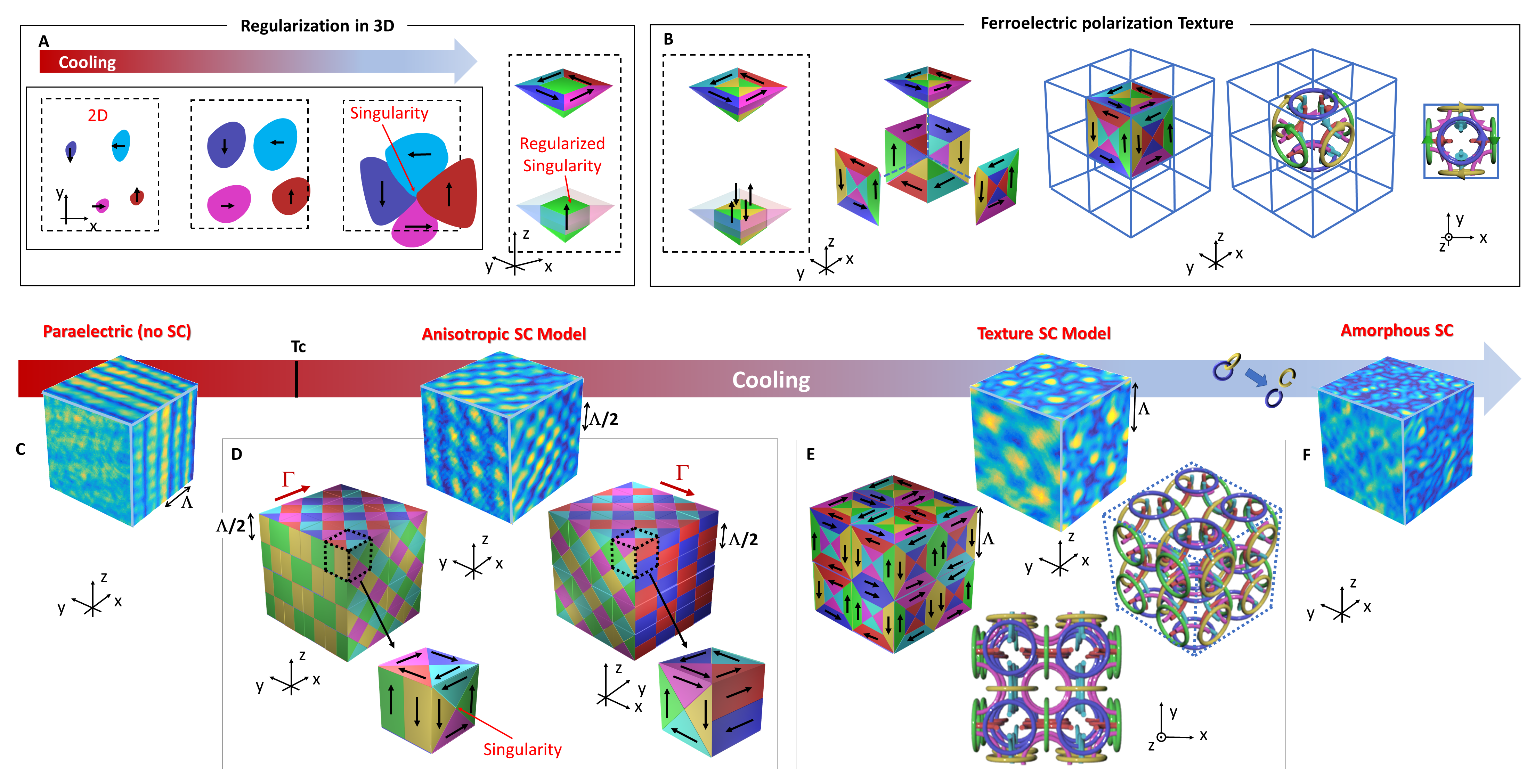}
\caption{Modelling the hyperdimensional topological defect formation.   (A) Kibble mechanism for 2D xy system (color coded, black arrows represent $\mathbf{P}$).  As the system is cooled to the critical temperature, initially independent domains (first panel)  grow (second panel) and can lock into a vortex-like singularity (third panel).  The singularity is regularized forming a z polarized domain. The result is a ferroelectric meron for the now 3D xyz system (fourth panel).   (B) Merons can combine to form singularities in 3D that  regularize by forming a ferroelectric polarization texture. Each meron is modified to support linked closed-flux polarization loops (first panel).  As six component modified merons meet (second panel), they form a cubic structure that can fill 3D space (third panel), where no singularity remains and all polarization loops are closed-flux (third and fourth panels).  (C-F) Rendering of 3D stereoscopic actual space images for the different crystal phases on cooling through $T_C$. (C) Paraelectric phase for $T>T_C$. (D)  Anisotropic SC phase for $T_C-2$K. Stereoscopic rendering (top panel) compared to the color-coded model as seen for two different perspectives. (E) Texture SC phase for $T_C-3$K. Stereoscopic rendering (top panel) compared to the Texture SC model (left panel) and the resulting Hopf links as illustrated for two perspectives (right and bottom panels). The polarization distribution forms localized 3D structures distributed on a 3D periodic lattice. (F) Amorphous SC phase for $T_C-6$K after the Hopf links have been broken.}
\label{figure3}
\end{figure*}

The phenomenon is observed performing stereoscopic optical imaging of a 2.5 mm (x) by 2.0 mm (y) by 1.9 mm (z)   K$_{0.997}$Ta$_{0.64}$Nb$_{0.36}$O$_3$:Li$_{0.003}$ solid-solution on cooling through the room-temperature Curie point ($T_C=294$ K, see Supplemental Material for setup details) \cite{DelRe2011,Falsi2020,Presti2020}. Evidence of a texture supercrystal (SC) at $T=T_C-3$K is reported in Fig. 1 B-I.  In each panel, 3D wide-area  transmission images of the $xz$ sample facet are reported for various rotations (top row) and compared to what expected from the texture SC model (bottom row, model in inset).  Direct transmission through the SC with no tilt (Fig. 1 B) leads to a characteristic bcc (body-center-cubic) pattern with a $\Lambda \simeq 15\mu$m lattice constant.  As the 2 mm long ($y$-axis) sample is rotated in the horizontal plane, the interlocked-ring texture structure distorts the transmission peaks. This leads to the characteristic images reported in Fig. 1C,D for a rotation around the $x$-axis of $\theta_x \simeq 2$ mrad and $3$ mrad, respectively. Comparison is then shown to the texture 3D model rotated by the internal angle $\theta'=\theta/n$, appropriately rescaled to take refraction into account ($n\simeq 2.2$ (at 532nm), while larger and even giant refraction is observed in the sample for finite angles (not reported here) \cite{DiMei2018}).   In Fig. 1 E, a (relatively large) horizontal rotation of 8 mrad now causes direct transmission through the texture lattice cores to cease.  This allows the surrounding ring-like polarization structure at the output facet to scatter light and become visible.  An analogous set of transmission and scattering experiments emerge for rotations around the second $z$ axis, while a telling stereoscopy appears when the sample is rotated both around the $x$ and the $z$ axes, as reported in Fig. 1 F for $(\theta_x,\theta_z) \simeq$ (3 mrad, 1 mrad).  Fig.1 G, H and I report scattering off the output facet texture details for $(\theta_x,\theta_z) \simeq$ (3 mrad,  3 mrad), $(\theta_x,\theta_z) \simeq$ (7 mrad, 7 mrad), and $(\theta_x,\theta_z) \simeq$ (8 mrad,  8 mrad), respectively, in conditions where direct transmission ceases.  This reveals features that were not illuminated in Fig. 1 E and completes the overall imaging of the structure.   Analogous results are found analyzing transmission through the other facets (i.e., the $xy$ and $yz$ facets). The wide-area wide-angle snapshots also contain different perspectives of the extended Hopf-links in different regions, a rich phenomenology that can be appreciated in an experimental exploration reported in the Supplemental Video.

In Fig. 2 we report the imaging of the $yz$ (Fig. 2 A), $xz$ (Fig. 2 B), and $xy$ (Fig. 2 C) external facets of the sample on cooling to different temperatures.  For $T>T_C$, an image of the output facet of the sample shows a characteristic striation pattern, a quasi-periodic planar index of refraction distribution, with the planes oriented parallel to $\mathbf{\Gamma}$ (first column panels), i.e., the direction along which the sample is slowly pulled as it deposits from the melt.  At $T=T_C-2$K, while  the images of  the $xz$ and $xy$ facets are compatible with a bcc structure, the image of the $yz$ facet (second column) is compatible with a  simple-cubic structure, an anisotropic signature that is in agreement with previous literature \cite{Pierangeli2016}. The lattice constant of the pattern is $\Lambda/2$=7.5~$\mu$m. As the sample is cooled to the texture SC state analyzed in Fig. 1 ($T=T_C-3$K), 3D sterescopic direct imaging (third column) and far-field imaging of the facets (see Fig. S2 in Supplementary Material) indicates an authentic bcc stucture with a lattice constant $\Lambda$ ($\simeq 15\mu$m), the highly symmetric distribution expected for a simultaneous inversion-symmetry-breaking in all three spatial directions. For even lower temperatures (fourth column panels), direct imaging leads to a strongly anisotropic amorphous distribution.  As reported in Fig. 2 D, imaging the Hopf link structure  (as in Fig. 1 E) suggests that the passage from the texture SC to the amorphous phase involves the breaking of the single links on cooldown,  a characteristic feature of Kibble topological defects \cite{Kosterlitz2017}. 

To analyze time-inversion symmetry, we compared sample stereoscopic imaging for cooling and heating through the Curie point.  Results reported in Fig. 2 E  demonstrate that the SC texture phase depends on thermal history  (black squares, cooling, red squares, heating) as seen through all three  $yz$, $xz$, and $xy$ facets (top, center, and bottom panels, respectively).  The finding is in agreement with the Kibble mechanism: textures occur when the breaking of the three-fold spatial inversion symmetry (the formation of 3D ferroelectric cluster patterns) is accompanied by the breaking of the one-fold time-inversion-symmetry (thermal hysteresis).

Results can be discussed starting from the explanatory model of ferroelectric supercrystals (SC), a mosaic of ferroelectric domains that can form as a perovskite undergoes its ferroelectric phase-transition \cite{Pierangeli2016,FERRARO2017,Zhang2019,Yang2020,Falsi2020,Hadjimichael2021}.  Consider the formation of ferroelectric topological defects arising from planar-polarized  (2D) domains, that is, where the spontaneous polarization field at each point in space $\mathbf{r}_i$ is a constant-amplitude field oriented only parallel or antiparallel to the two principal axes, $\mathbf{P(\mathbf{r}_i)} \in \{\pm P\mathbf{\hat{x}}, \pm P\mathbf{\hat{y}}\}$,  (xy-discrete-strain-model) \cite{Muench2019}.  The Kibble mechanism, illustrated in Fig. 3 A, begins as  spontaneous polarization (black arrows)  $\mathbf{P}$ enucleates independent domains on relaxation below the critical temperature $T_C$ (from left to right panels) \cite{Forsbergh1949,Bokov2006,Roytburd2017}.  In the presence of multiple clusters, domain-domain interaction is dominated by the reduction of volume polarization charge and strain, a condition that can spontaneously generate characteristic  closed-loop polarization distributions. As the system relaxes, the initially independent polarized domains (first panel) expand and interact (second panel), ultimately locking into a stalemate around a polarization vortex (third panel) \cite{Naumov2004,Stoica2019,Das2019}.  The core of the vortex is not polarized in the 2D layer, forming a singularity whose regularization occurs as the polarization field escapes into the third dimension (fourth panel).  The vortex core forms now an extended string that is in fact a polarized domain in the third $z$ direction, the polarization field  having a full 3D structure $\mathbf{P(\mathbf{r}_i)} \in \{\pm P\mathbf{\hat{x}}, \pm P\mathbf{\hat{y}}, \pm P\mathbf{\hat{z}} \}$,  (xyz-discrete-strain-model for tetragonal perovskites \cite{Muench2019}). The regularized structure is a meron, the half-component of a skyrmion \cite{Lin2015,Wang2020}.  In the same manner, the Kibble mechanism generates singularities as different independent merons meet, meron composites that are then regularized as the polarization field forms interlinked closed loops in a full globally-interlinked Hopf fibration (Fig. 3 B).  As for the passage from 2D to 3D, the regularization requires the polarization field to escape into the fourth dimension, i.e., breaking also the time-reversal symmetry.

In these terms, the paraelectric state observed at $T>T_C$ (Fig. 3C) can be associated to temperatures for which the ferroelectric clusters are non-interacting.  In turn, the transient phase we term Anisotropic can be associated to the transient formation illustrated in Fig. 3 D. This state is superseded by the stable Hopf-fibration lattice state illustrated in Fig. 3E (Fig. 1 and Fig. 2 A, B, and C third and fourth column), with its characteristic breaking of time-reversal symmetry (Fig. 2 E). The Hopf fibration, that corresponds to a superlattice of polarization hyperdimensional textures,  ultimately dissolves as the rings breakup at even lower temperatures, as illustrated in Fig. 3 F (Fig. 2 D).

We note that topological defects can also be artificially induced  by boundary conditions and external constraints     \cite{Mermin1979,Chuang1991,Bowick1994,Hasan2010,Nagaosa2013}. These can form Hopf-like knots, such as Hopfions in liquid-crystals \cite{Chuang1991,Chen2013,Gim2017,Tai2019}, ferromagnetic colloidal suspensions \cite{Ackerman2017}, and ferroelectric nanoparticles \cite{Lukyanchuk2020}. Attached to a boundary (the surface of the film, nanoparticle, or colloidal particle), these do not form the spatially extended and globally-interlinked structure that is expected of a Kibble texture.

We have carried out three-axis stereoscopic imaging of a KTN:Li crystal undergoing its room-temperature ferroelectric phase-transition.  Results  indicate the existence of a ferroelectric cluster phase below the Curie point that appears as a stable lattice of interlinked hyperdimensional $N=4$ texture topological defects arising from the simultaneous breaking of a three-axis spatial-inversion and time-inversion symmetry.  Our demonstration of texture formation provides a first answer to the question of how spontaneous extra-dimensional wrap-up can form.   Arising during a phase transition through the universal Kibble mechanism, our findings shed light on the nature of extra dimensions in other systems undergoing symmetry-breaking transitions, including ferromagnetic materials \cite{Ackerman2017}, superconductors \cite{Blatter1994}, and in the wider context of cosmology and the early universe \cite{Kolb1981,Peacock1998}. From an applicative perspective, the study can suggest new full 3D schemes in non-volative ferroelectric memories and negative-capacitance super-fast electronics \cite{Naumov2004,Das2019,Yadav2019}.

\section*{Acknowledgements}

We acknowledge support from the Attract Consortium (SALT project, H2020), Sapienza-Ricerca di Ateneo 2019 and 2020 projects, the H2020 Fet project PhoQus,  the PRIN 2017 PELM (Grant No. 20177PSCKT), the Israel Science Foundation (Grant No. 1960/16), and National Natural Science Foundation of China (Grant No.12004282).


\begin{thebibliography}{25}


\bibitem{Strocchi2008} F. Strocchi, \textit{Symmetry Breaking}, Lect. Notes Phys. 732 (Springer, Berlin Heidelberg 2008).

\bibitem{Kibble1976}  T. W. B. Kibble, ``Topology of cosmic domains and strings," \emph{J. Phys. A: Math. Gen.} \textbf{9} (8): 1387–1398 (1976).



\bibitem{Vilenkin1994}  A. Vilenkin, \& E. P. S. Shallard,  \textit{Cosmic strings and other topological defects} (Cambridge University Press, 1994). 



\bibitem{Griffin2012} S. M. Griffin, et al. ``Scaling Behavior and Beyond Equilibrium in the Hexagonal Manganites," \emph{Phys. Rev. X} \textbf{2}, 041022 (2012). 


\bibitem{Zong2018} A. Zong, et al. ``Evidence for topological defects in a photoinduced phase transition," \emph{Nature Phys.} \textbf{15}, 27–31 (2019). 

\bibitem{Bunkov2000}  Y. M. Bunkov, et al. \textit{Topological Defects and the Non-Equilibrium Dynamics of Symmetry Breaking Phase Transitions} (Kluwer Academic Publishers 2000). 



\bibitem{Lin2014} S.-Z. Lin, et al. ``Topological defects as relics of emergent continuous symmetry and Higgs condensation of disorder in ferroelectrics," \emph{Nat. Phys.} \textbf{10}, 970-977 (2014). 
%


\bibitem{Kosterlitz2017}  J. M. Kosterlitz, ``Nobel Lecture: Topological defects and phase transitions," \emph{Rev. Mod. Phys.} \textbf{89}, 040501 (2017). 

\bibitem{Tetienne2015} J. Tetienne, et al. ``The nature of domain walls in ultrathin ferromagnets revealed by scanning nanomagnetometry," \emph{Nat. Commun.} \textbf{6}, 6733 (2015).

\bibitem{Yadav2016}  A. Yadav, et al. ``Observation of polar vortices in oxide superlattices," \emph{Nature} \textbf{530}, 198–201 (2016).

\bibitem{Naumov2004} I. I. Naumov, L. Bellaiche, and H. Fu, ``Unusual phase transitions in ferroelectric nanodisks and nanorods," \emph{Nature} \textbf{432}, 737 (2004).


\bibitem{Hong2017} Z. Hong, et al. ``Stability of Polar Vortex Lattice in Ferroelectric Superlattices," \emph{Nano Lett.} \textbf{17}, 2246-2252 (2017).

\bibitem{Hsu2019}  S. Hsu, M. R. McCarter, C. Dai, Z. Hong, L. Chen, C. T. Nelson, L. W. Martin, and R. Ramesh,
``Emergence of the Vortex State in Confined Ferroelectric Heterostructures," \emph{Advanced Materials}, \textbf{31}, 1901014 (2019).




\bibitem{Nagaosa2013}  N. Nagaosa, and Y. Tokura, ``Topological properties and dynamics of magnetic skyrmions," \emph{Nature Nanotech.} \textbf{8}, 899–911 (2013).


\bibitem{Yu2018} X. Z. Yu, et al. ``Transformation between meron and skyrmion topological spin textures in a chiral magnet," \emph{Nature} \textbf{564}, 95–98 (2018).

\bibitem{Posnjak2017} G. Posnjak, S. $\check{C}$opar,  I. \& Mu$\check{s}$evi$\check{c}$,  ``Hidden topological constellations and polyvalent charges in chiral nematic droplets," \emph{Nat. Commun.} \textbf{8}, 14594, (2017).


\bibitem{Kanazawa2020} N. Kanazawa et al. ``Direct Observation of the Statics and Dynamics of Emergent Magnetic Monopoles in a Chiral Magnet," \emph{Phys. Rev. Lett.} \textbf{125}, 137202 (2020)



\bibitem{Das2020} S. Das, et al. ``Local negative permittivity and topological phase transition in polar skyrmions," \emph{Nat. Mater.} (2020). 


\bibitem{Das2019} S. Das, et al. ``Observation of room-temperature polar skyrmions," \emph{Nature} \textbf{568}, 368-372 (2019).

\bibitem{Note1}  While a texture in cosmology is an $N=4$ hyperdimensional topological defect distinct from a wall, vortex, or skyrmion \cite{Kibble1976,Chuang1991,Vilenkin1994}, the term has also been recently used to refer to patterns (``spin textures'') of merons and skyrmions  \cite{Yu2018,Das2020}.



\bibitem{Derrick1964} G. H. Derrick, ``Comments on Nonlinear Wave Equations as Models for Elementary Particles," \emph{Journal of Mathematical Physics} \textbf{5}, 1252-1254 (1964).

\bibitem{DelRe2011}  E. DelRe, E. Spinozzi,  A. J. Agranat, and C. Conti, 
``Scale-free optics and diffractionless waves in nanodisordered ferroelectrics," \emph{Nat. Photon.} \textbf{5}, 39-42 (2011)

\bibitem{Falsi2020} L.  Falsi, et al. ``Constraint-free wavelength conversion supported by giant optical refraction in a 3D perovskite supercrystal," \emph{Communications Materials} \textbf{1}, 76 (2020).


\bibitem{Presti2020} L. L. Presti, et al. ``Observation of an exotic lattice structure in the transparent KTa$_{1-x}$Nb$_x$O$_3$ perovskite supercrystal," \emph{Phys. Rev. B} \textbf{102}, 214110 (2020).


\bibitem{DiMei2018} F. Di Mei, L. Falsi, M. Flammini, D. Pierangeli, P. Di Porto, A. J. Agranat, and E. DelRe, ``Giant broadband refraction in the visible in a ferroelectric perovskite," \emph{Nat. Photon.} \textbf{12}, 734 (2018).

\bibitem{Pierangeli2016} D. Pierangeli, M. Ferraro, F. Di Mei, G. Di Domenico, C. E. M. de Oliveira, A. J. Agranat, and E. DelRe,  ``Super-crystals in composite ferroelectrics,'' \emph{Nat. Commun.} \textbf{7}, 10674 (2016).


\bibitem{FERRARO2017} M. Ferraro, et al. ``Observation of polarization-maintaining light propagation in depoled compositionally disordered ferroelectrics," \emph{Opt. Lett.} 42, 3856-3859 (2017).

\bibitem{Zhang2019} X. Zhang, et al.
``Switching effects of spontaneously formed superlattices in relaxor ferroelectrics," \emph{Opt. Mat. Express} \textbf{9}, 4081-4089 (2019).

\bibitem{Yang2020} Q. Yang, et al. ``Dynamic relaxation process of a 3D super crystal structure in a Cu:KTN crystal," \emph{Chin. Opt. Lett.} \textbf{18}, 021901(2020).

\bibitem{Hadjimichael2021}  M. Hadjimichael,et al. ``Metal–ferroelectric supercrystals with periodically curved metallic layers," \emph{Nat. Mater.} (2021).

\bibitem{Muench2019} I. Muench, A. Renuka Balakrishna, J. E. Huber, ``Periodic boundary conditions for the simulation of 3D domain patterns in tetragonal ferroelectric material", \emph{Arch Appl Mech} \textbf{89}, 955–972 (2019).












\bibitem{Forsbergh1949} P. W. Forsbergh, ``Domain Structures and Phase Transitions in Barittm Titanate," \emph{Phys. Rev.} \textbf{76}, 1187 (1949).

\bibitem{Bokov2006}  A. Bokov,  ``Recent progress in relaxor ferroelectrics with perovskite structure," \emph{J. Mater. Sci.} \textbf{41}, 31 (2006). 

\bibitem{Roytburd2017} Roytburd, A. L., Ouyang, J., and A. Artemev,   ``Polydomain structures in ferroelectric and ferroelastic epitaxial films,"  \emph{J. Phys.: Condens. Matter} \textbf{29}, 163001 (2017).

\bibitem{Stoica2019} V. A. Stoica, N. Laanait, C. Dai, et al.,   ``Optical creation of a supercrystal with three-dimensional nanoscale periodicity,'' \emph{Nat. Mater.} \textbf{18}, 377-383 (2019). 
 
\bibitem{Lin2015}  S. Lin, A. Saxena, and C. D. Batista, ``Skyrmion fractionalization and merons in chiral magnets with easy-plane anisotropy,"
\emph{Phys. Rev. B} 91, 224407 (2015).


\bibitem{Wang2020}  Y.J. Wang,et al. ``Polar meron lattice in strained oxide ferroelectrics," \emph{Nat. Mater.} \textbf{19}, 881–886 (2020).

\bibitem{Mermin1979} N. D. Merminf , and H. Wagner, ``Absence of Ferromagnetism or Antiferromagnetism in One- or Two-Dimensional Isotropic Heisenberg Models," \emph{Phys. Rev. Lett.} \textbf{17}, 1307 (1966).

\bibitem{Hasan2010}  M. Z. Hasan, and C. L. Kane, ``Colloquium: Topological insulators," \emph{Rev. Mod. Phys.} \textbf{82}, 3045 (2010).

\bibitem{Chuang1991}  I. Chuang,  R. Durrer, N. Turok, and B. Yurke, ``Cosmology in Defect Dynamics the Laboratory: in Liquid Crystals," \emph{Science} \textbf{251}, 1336-1342 (1991).

\bibitem{Bowick1994} M. J. Bowick, L. Chandar, E. A. Schiff, and  A. M. Srivastava, ``The Cosmological Kibble Mechanism in the Laboratory: String Formation in Liquid Crystals," \emph{Science} \textbf{263}, 943-945 (1994).



\bibitem{Chen2013} B. G. Chen, et al. ``Generating the Hopf Fibration Experimentally in Nematic Liquid Crystals," \emph{Phys. Rev. Lett.} \textbf{110}, 237801 (2013).


\bibitem{Tai2019} J. B. Tai, and I. I. Smalyukh, ``Three-dimensional crystals of adaptive knots," \emph{Science} \textbf{365}, 1449-1453 (2019).

\bibitem{Gim2017}  M. J. Gim, D. Beller, and D. Yoon, ``Morphogenesis of liquid crystal topological defects during the nematic-smectic A phase transition," \emph{Nat. Commun.} \textbf{8}, 15453 (2017).

\bibitem{Ackerman2017}  P. J. Ackerman, and I. I. Smalyukh, ``Static three-dimensional topological solitons in fluid chiral ferromagnets and colloids," \emph{Nat. Mater.} \textbf{16}, 426-433 (2017).

\bibitem{Lukyanchuk2020}  I. Luk'yanchuk, Y. Tikhonov, A. Razumnaya, and  V. M. Vinoku, ``Hopfifions emerge in ferroelectrics," \emph{Nat. Commun.} \textbf{11}, 2433 (2020).








\bibitem{Blatter1994} G. Blatter,  M. V. Feigel'man, V. B. Geshkenbein,  and A. I. Larkin, ``Vortices in high-temperature superconductors," \emph{Rev. Mod. Phys.} \textbf{66}, 1125–1388 (1994).

\bibitem{Kolb1981} E. Kolb, M. Turner,  ``The early Universe," \emph{Nature} \textbf{294}, 521–526 (1981).

\bibitem{Peacock1998} J. Peacock, \textit{Cosmological Physics} (Cambridge University Press, 1998).

\bibitem{Yadav2019} A. K. Yadav,  et al.,`` Spatially resolved steady-state negative capacitance," \emph{Nature} \textbf{565}, 468–471 (2019). 

\end{thebibliography}
\end{document}